\documentclass[ 11pt]{article}
\usepackage{graphicx,amsmath,amssymb,epsf}
\usepackage{latexsym,tabularx,bm, cite, verbatim, fdag}
\newcommand{\tr}{{\rm tr}} 

\begin{document}

\title{ \Large \bf Pole prescription of higher order induced vertices in Lipatov's QCD   effective action}
\author{\large  Martin~Hentschinski
\bigskip  \\
 Instituto de F\'isica Te\'orica UAM/CSIC, \\  Universidad Aut\'onoma de Madrid, \\ Cantoblanco,  E-28049 Madrid, Spain
}

\maketitle

\vspace{-9cm}
\begin{flushright}
  IFT-UAM/CSIC-11-105 \\
LPN11-95 
\end{flushright}
\vspace{6.5cm}

\begin{abstract}
  We investigate Lipatov's QCD effective action for the QCD high
  energy limit and propose a pole prescription for higher order
  induced vertices. The latter can be used in the evaluation of loop
  corrections to high energy factorized matrix elements within the
  effective action approach.  The proposed prescription respects the
  symmetry properties of the unregulated vertices. Explicit results
  are presented up to third order in the gauge coupling, while an
  iterative procedure for higher orders is proposed.
\end{abstract}

\section{Introduction}
\label{sec:intro}

The description of the high energy limit of perturbative QCD is
generally given in terms of high energy factorization and the BFKL
evolution. The latter resums large logarithms in the center of mass
energy $\sqrt{s}$ at leading (LL)
\cite{Fadin:1975cb}
and next-to-leading logarithmic  (NLL) accuracy \cite{Fadin:1998py}. Its
derivation is closely related to the observation that QCD scattering
amplitudes reveal in the high energy limit an effective $t$-channel
degree of freedom, the reggeized gluon, which couples to the external
scattering particles through effective vertices.  Determination of
both  effective couplings and higher order corrections to 
reggeized gluon exchange is a highly  non-trivial task. This
is especially true if one attempts to go beyond leading order accuracy
-- something which is needed  for {\it e.g.} a successful
phenomenology of the QCD high energy limit.

An efficient tool to address these questions is given by Lipatov's
high energy effective action \cite{Lipatov:1995pn}.
It is based on the QCD action with the addition of an induced
term. The latter is written in terms of gauge-invariant currents which
generate a non-trivial coupling of the gluon to reggeized gluon
fields.  The effective action allows to address both unitarization of
BFKL evolution and the systematic determination of higher order
perturbative corrections to high energy QCD amplitudes. While the high
energy limit of QCD tree-level amplitudes can be obtained in a
straightforward manner from the effective action
\cite{Antonov:2004hh}, in loop corrections a new type of longitudinal
divergences, not present in conventional QCD amplitudes, appear.  The
treatment of these divergences has been at first addressed for
leading order transition kernels
\cite{Hentschinski:2008rw,Hentschinski:2008im} and the study of the leading order reggeized
gluon - gluon - 2 reggeized gluon (RG2R) production vertex in
\cite{Braun:2011it,  Braun:2006sk}.  Recently the
effective action has been used for the first time for the calculation
of NLO corrections to the forward quark jet impact factor
\cite{Hentschinski:2011tz} and the quark part of the 2-loop gluon
trajectory \cite{Chachamis:prep}, finding precise agreement with
previous results in the literature.

In these studies of loop corrections it has been further necessary to
give a prescription for circumventing light-cone singularities on the
complex plane. The latter appear due to a non-local operator in the
induced term of the effective action which describes the coupling of
reggeized gluons to conventional gluons.  Locality of the effective
action in rapidity space and high energy kinematics prevent in
principle the light-cone momenta in these denominators to take
non-zero values. A prescription for these poles seems not to be
necessary. That this is true is evident in the case of tree-level
amplitudes in the (Quasi-)Multi-Regge-Kinematics. It seems furthermore
reasonable to expect that a similar statement holds for corresponding
virtual corrections.  On the other hand all regularizations used up
till now need a prescription for these denominators. Other
regularizations which contain more physical insight are surely
possible, but remain to be explored. In this work we therefore give a
pole prescription which can be applied in combination with
regularization methods used so far.

To leading order (LO)  in the gauge coupling $g$, the order $g$ induced
vertex has so far been interpreted as a Cauchy principal value
\cite{Hentschinski:2008rw, Hentschinski:2008im, Chachamis:prep,
  Hentschinski:2011tz, Braun:2006sk,Braun:2011it, Bartels:prep}.
Higher order induced vertices seem to require a more
elaborate prescription. In particular interpreting n\"aively every
single pole of the higher vertices as a Cauchy principal value
destroys parts of the symmetry of the unregulated vertices and can
lead to incorrect results, see for instance
\cite{Hentschinski:2008rw}.

In the following we provide a pole prescription of higher order
induced vertices which respects the symmetry properties of the
unregulated vertices and which can be motivated through the high
energy expansion of QCD scattering amplitudes.  We will present
explicit results up to the order $g^3$ induced vertex which can be
directly used for calculations within the effective action such as the
determination of the gluon part of the 2-loop gluon Regge trajectory,
corrections to gluon induced production processes and studies of
multiple reggeized gluon exchanges within the effective action.

The paper is organized as follows: in sec.~\ref{sec:lagrangian} we
give a short introduction to Lipatov's effective action, including a
summary of unregulated Feynman rules. In sec.~\ref{sec:poles} we
present a derivation of the proposed pole prescription and provide
explicit expressions up to third order in the gauge coupling. Finally
in sec.~\ref{sec:concl} conclusions and suggestions for future work
are presented. The appendix contains a simple QCD example which
illustrates the relation of the chosen pole prescription with
underlying QCD amplitudes.

\section{The effective action of high energy QCD}
\label{sec:lagrangian}
\begin{figure}[htb]
  \centering
   \parbox{3cm}{\includegraphics[height = 2cm]{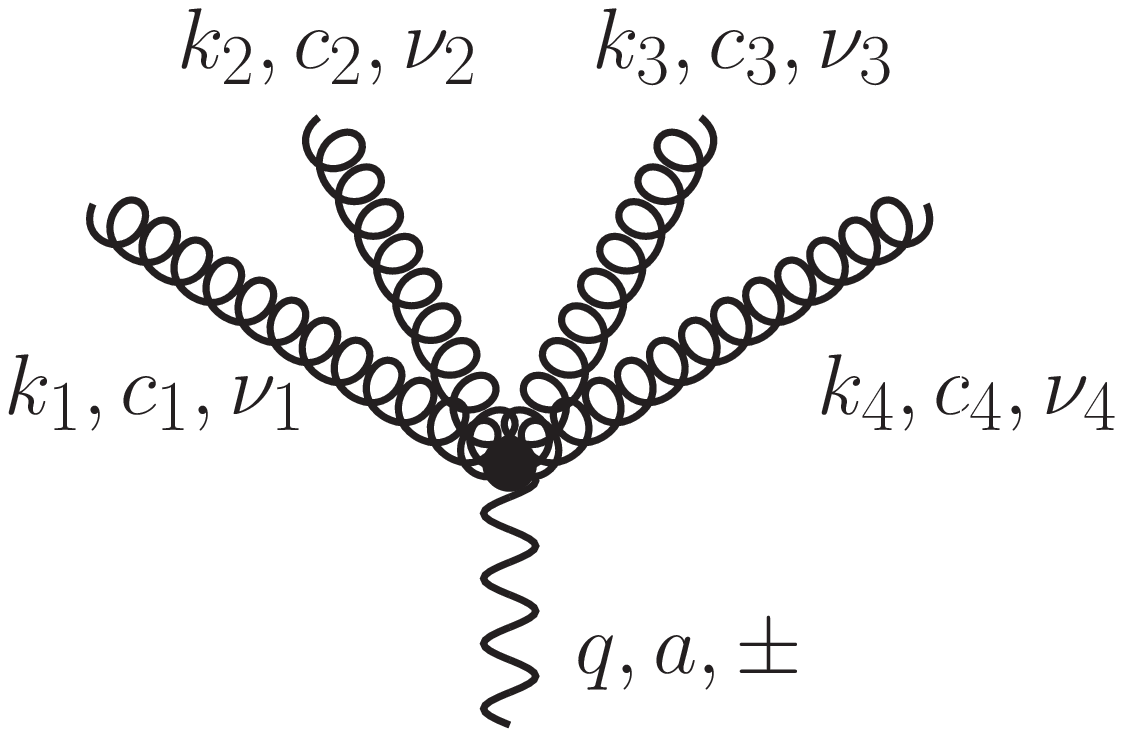}}  $ \begin{array}[h]{ll}  \\  =
   \displaystyle 
&   \displaystyle   g^3 {\bm{q}}^2  
\bigg[
 \frac{f^{c_4c_3 e_2}}{k_4^\pm}
\bigg(
\frac{f^{e_2c_1e_1} f^{c_2e_1a}}{(k_1^\pm + k_2^\pm)k_2^\pm} +  \frac{f^{e_2c_2e_1} f^{c_1e_1a}}{(k_1^\pm + k_2^\pm)k_1^\pm}
\bigg)  \\ \\ \displaystyle 
& \displaystyle  + \,
\frac{f^{c_4c_1 e_2}}{k_4^\pm}
\bigg(
\frac{f^{e_2a_2e_1} f^{c_3e_1a}}{(k_3^\pm + k_2^\pm)k_3^\pm} +
 \frac{f^{e_2a_3e_1} f^{c_2e_1a}}{(k_3^\pm + k_2^\pm)k_2^\pm}
\bigg)  \, + \\ \\
   \end{array} $ \\
$
\begin{array}[h]{l}
  \displaystyle + 
\frac{f^{c_4c_2 e_2}}{k_4^\pm}
\bigg(
\frac{f^{e_2c_1e_1} f^{c_3e_1a}}{(k_3^\pm + k_1^\pm)k_2^\pm} +  \frac{f^{e_2c_2e_1} f^{c_1e_1a}}{(k_3^\pm + k_1^\pm)k_3^\pm}
\bigg)
\bigg](n^\pm)^{\nu_1} (n^\pm)^{\nu_2}  (n^\pm)^{\nu_3}  (n^\pm)^{\nu_4}  \\ \\ \qquad \qquad  \qquad \qquad \qquad \qquad  \qquad k_1^\pm  + k_2^\pm +  k_3^\pm  + k_4^\pm = 0
\end{array}
$
  \caption{\small The unregulated order $g^3$ induced vertex }
  \label{fig:trans3}
\end{figure}

To set the notation used in the following for the effective action, it
is useful to have in mind a partonic elastic scattering process $p_a +
p_b \to p_1 + p_2$ with $p_a^2 = 0 = p_b^2$ light-like momenta and $ s = (p_a + p_b)^2 =
2 p_a\cdot p_b$ the squared center of mass energy. We then define
light-like four vectors $n^\pm$ with $n^+\cdot n^- = 2$, related to
the momenta of scattering partons by the re-scaling, $ n^+ = {2 p_b
}/{\sqrt{s}} $ and $ n^- = {2 p_a}/{\sqrt{s}} $.  The Sudakov
decomposition of a general four vector $k^\mu$  reads $ k = {k^+}
n^-/{2} + {k^-} n^+/{2} + {\bm k} $, where $k^\pm = n^\pm\cdot k $ and
${\bm k}$ is transverse w.r.t. the initial scattering axis.  The
effective action is given as the sum of two terms, $S_{\text{eff}} = S_{\text{QCD}} + S_{\text{ind.}}$, the QCD action and the induced term. The latter  describes
the coupling of the reggeized gluon field $A_\pm(x) = -i t^a A_\pm^a(x)
$  to the gluonic field $v_\mu(x) = -it^a v_\mu^a(x)$. It
reads
\begin{align}
\label{eq:1efflagrangian}
  S_{\text{ind.}}[v_\mu, A_\pm]& = \int \! \text{d}^4 x \,
\tr\bigg[\bigg( W_+[v(x)] - A_+(x) \bigg)\partial^2_\perp A_-(x)\bigg]
\notag \\
&  \qquad  \qquad   \qquad
+\int \! \text{d}^4 x \,\tr\bigg[\bigg(W_-[v(x)] - A_-(x) \bigg)\partial^2_\perp A_+(x)\bigg]
.
\end{align}
The infinite number of couplings of the gluon field to the reggeized
gluon field are contained in two functionals $W_\pm[v] $, which are
defined through the following operator definition
\begin{align}
\label{eq2:efflagrangian}
W_\pm[v] =
&
v_\pm \frac{1}{ D_\pm}\partial_\pm  && \text{where} & D_\pm &= \partial_\pm + g v_\pm .
 \end{align}
  For perturbative calculations,  the
 following  expansion in the gauge coupling $g$ holds, 
\begin{align}
  \label{eq:funct_expand}
  W_\pm[v] =&  v_\pm - g  v_\pm\frac{1}{\partial_\pm} v_\pm + g^2 v_\pm
\frac{1}{\partial_\pm} v_\pm\frac{1}{\partial_\pm} v_\pm - \ldots
\end{align}
The determination of a suitable regularization of the above operators
$1/\partial_\pm$ at zero is then the main goal of this work. Note
that the reggeized gluon fields are special in the sense that they are
invariant under local gauge transformations, while they transform
globally in the adjoint representation of the SU$(N_c)$ gauge
group. In addition strong ordering of longitudinal momenta in high
energy factorized amplitudes leads to the following kinematical
constraint of the reggeized gluon fields,
\begin{align}
  \label{eq:constraint}
\partial_+ A_-(x)  =  \partial_- A_+ (x)= 0,
\end{align} 
which is always implied. 
\begin{figure}[htb]
  \centering
   \parbox{.7cm}{\includegraphics[height = 1.8cm]{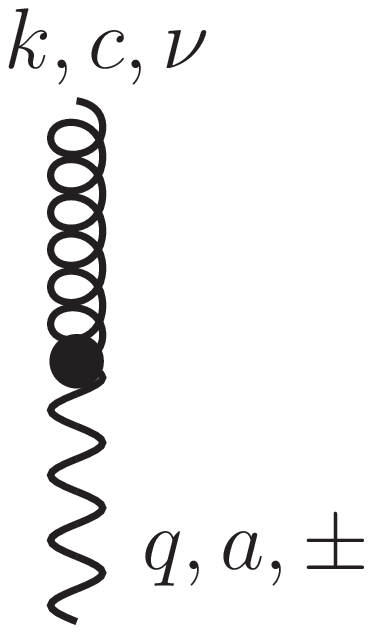}} $=  \displaystyle 
   \begin{array}[h]{ll}
    \\  \\ - i{\bm q}^2 \delta^{a c} (n^\pm)^\nu,  \\ \\  \qquad   k^\pm = 0.
   \end{array}  $ 
 \parbox{1.2cm}{ \includegraphics[height = 1.8cm]{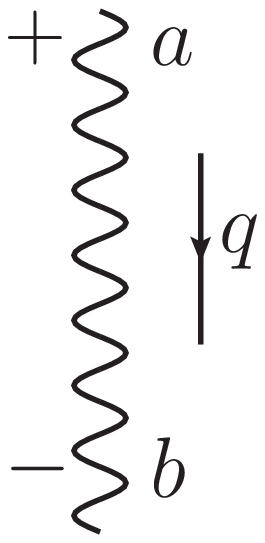}}  $=  \displaystyle    \begin{array}[h]{ll}
    \delta^{ab} \frac{ i/2}{{\bm q}^2} \end{array}$ 
 \parbox{1.7cm}{\includegraphics[height = 1.8cm]{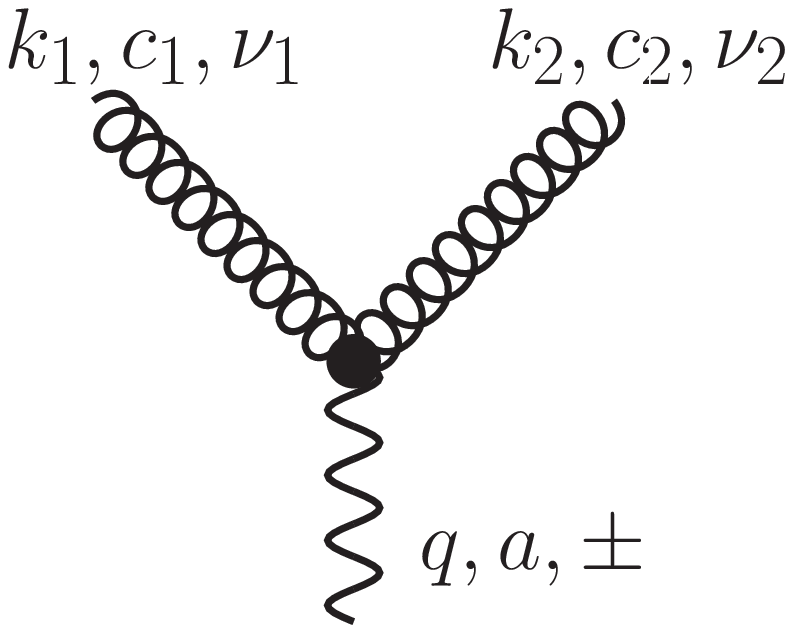}} $ \displaystyle  =  \begin{array}[h]{ll}  \\ \\ g f^{c_1 c_2 a} \frac{{\bm q}^2}{k_1^\pm}   (n^\pm)^{\nu_1} (n^\pm)^{\nu_2},  \\ \\ \quad  k_1^\pm  + k_2^\pm  = 0
 \end{array}$
 \\
\parbox{4cm}{\center (a)} \parbox{4cm}{\center (b)} \parbox{4cm}{\center (c)}
  \caption{\small The direct transition vertex (a), the reggeized gluon propagator (b) and the unregulated order $g$ induced vertex (c) }
  \label{fig:feynrules0p2}
\end{figure}
Feynman rules of the high energy effective action have been determined
in \cite{Antonov:2004hh}. We depict them in the following using curly
lines for the conventional QCD gluon field and wavy (photon-like)
lines for the reggeized gluon field.  Following the convention of
\cite{Antonov:2004hh}, the Feynman rules of the effective action are
given by the conventional QCD Feynman rules and an infinity number of
induced vertices, which include a direct transition term between gluon
and reggeized gluon, fig.~\ref{fig:feynrules0p2}.a. Within this direct
transition picture, the propagator of the reggeized gluon receives at
tree-level a correction due to the projection of the gluon propagator
\begin{figure}[htb]
  \centering
   \parbox{2.4cm}{\includegraphics[height = 1.8cm]{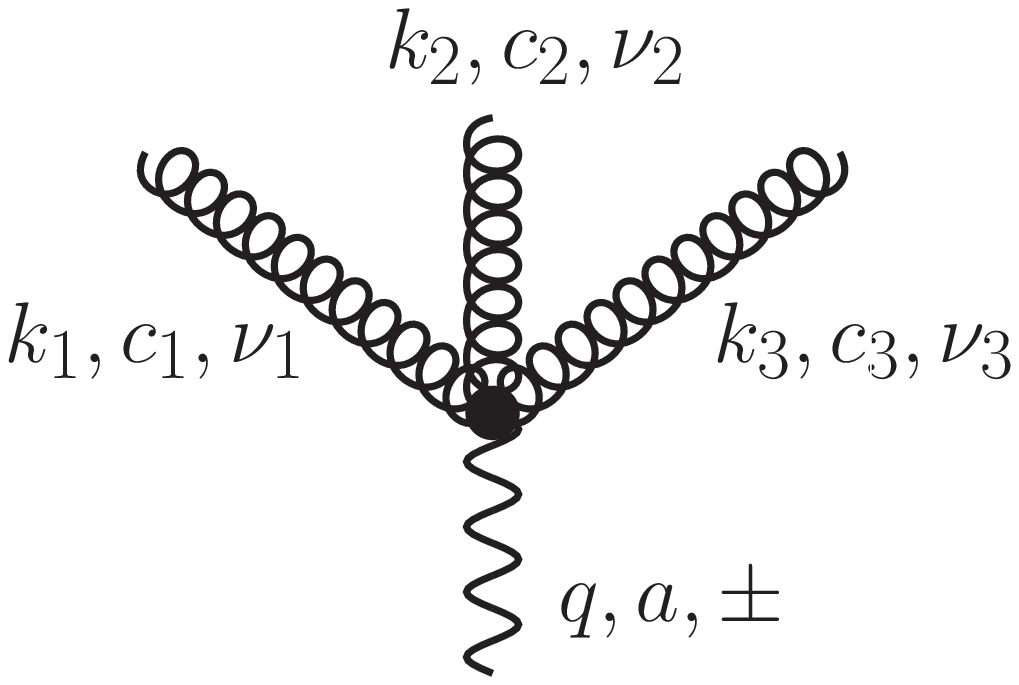}} $ \displaystyle 
   \begin{array}[h]{l}  \displaystyle  \\ \displaystyle= ig^2 {\bm{q}}^2 
\left(\frac{f^{a_3a_2 e} f^{a_1ea}}{k_3^\pm k_1^\pm} 
+
 \frac{f^{a_3a_1 e} f^{a_2ea}}{k_3^\pm k_2^\pm}\right) (n^\pm)^{\nu_1} (n^\pm)^{\nu_2} (n^\pm)^{\nu_3} \\ \\
\qquad \qquad   k_1^\pm + k_2^\pm + k_3^\pm = 0  
   \end{array}
$
  \caption{\small  The unregulated order $g^2$ induced vertex}
  \label{fig:tra2}
\end{figure}
on the reggeized gluon states which we absorb here into the bare
reggeized gluon propagator, fig.~\ref{fig:feynrules0p2}.b. Higher
order reggeized gluon - $n$ gluon transition vertices are up to
$\mathcal{O}(g^3)$ given by figs.~\ref{fig:feynrules0p2}.c,
\ref{fig:tra2} and \ref{fig:trans3}.  Note that there exists a general
iterative formula for the $\mathcal{O}(g^n)$ vertex which is contained
in \cite{Antonov:2004hh}.  All of these vertices obey Bose-symmetry,
{\it i.e.}  symmetry under simultaneous exchange of color,
polarization and momenta of the external gluons of the order $g^n$
vertex. This can be verified making use of the constraint $
\sum_{i}^{n+1}k_i^\pm = 0$, which is a direct consequence from
eq.~\eqref{eq:constraint}, and the Jacobi identity.

\section{Pole prescription for induced vertices}
\label{sec:poles}
In its original formulation the effective action does not specify a
prescription for the poles of induced vertices.  A common choice
adapted to the induced vertex fig.~\ref{fig:feynrules0p2}.c is to
interpret the pole as a Cauchy principal value,
\cite{Braun:2006sk,Hentschinski:2008rw,Hentschinski:2008im,
  Hentschinski:2011tz}
\begin{align}
  \label{eq:indu1eps}
   \parbox{2.2cm}{\includegraphics[height = 2cm]{indu1.eps}} &  = g f^{c_1 c_2 a}   \frac{{\bm q}^2}{[k_1^\pm]}   (n^\pm)^{\nu_1} (n^\pm)^{\nu_2}, & \frac{1}{[k_1^\pm]} & \equiv   \frac{1}{2} \left( \frac{1}{k_1^\pm + i\epsilon }  + \frac{1}{k_1^\pm - i\epsilon}  \right).
\end{align}
This choice has the advantage that it maintains the symmetry of the
vertex without pole prescription; in particular both  Bose-symmetry
of the unregulated vertex and its anti-symmetry under the substitution
$k_1^\pm \to - k_1^{\pm}$ are kept. The latter property is of
importance as it can be directly related to negative signature of the
reggeized gluon, see \cite{Bartels:prep}. A straightforward extension
of the Cauchy principal value prescription to higher order vertices
interprets every separate pole as a Cauchy principal value.  However,
at least if n\"aively applied to the vertices in figs.~\ref{fig:tra2} and
\ref{fig:trans3}, such a prescription violates Bose-symmetry and can
lead to wrong results, see for instance \cite{Hentschinski:2008rw}.
This is mainly due to the fact that Cauchy principal values do not
obey the eikonal identity in an algebraic sense. One has instead an
additional term containing the product of two delta-functions,
\begin{align}
  \label{eq:eik_id_cpv}
\frac{1}{[k_1^\pm][k_1^\pm + k_2^\pm]}  + \frac{1}{[k_1^\pm][k_1^\pm + k_2^\pm]} = \frac{1}{[k_1^\pm][ k_2^\pm]} + \pi^2 \delta(k_1^\pm)\delta(k_2^\pm),
\end{align}
see also \cite{Bassetto:1991ue}.  In the following section we
 suggest a prescription which both respects
Bose-symmetry and negative signature of the reggeized gluon and
therefore maintains the symmetry properties of the unregulated
vertices.

\subsection{Pole prescription for order $g$ induced vertex}
\label{sec:poleind}

The most straightforward definition is to replace the operator $D_\pm$
in eq.~\eqref{eq:1efflagrangian} by its regulated counter part $D_\pm
- \epsilon$. This is a commonly used choice in other effective
theories with Wilson line like operators in the Lagrangian, see for
instance \cite{Beneke:2002ph}.  It has the advantage that it can be
related to the prescription which arises for these poles from an high
energy expansion of QCD diagrams. Up to the order $g^2$ induced vertex
this is explicitly demonstrated in Appendix \ref{sec:possig}, see also
the discussion in \cite{Balitsky:1995ub}.  For this regulation, the
regulated order $g$ induced vertex reads,
\begin{align}
 \label{eq:indu_1L_eps}
\left[ \parbox{2.4cm}{\includegraphics[height = 2cm]{indu1.eps}}\,\right]' & =   ig{\bm q}^2 2 \left( \frac{\tr(t^{c_1} t^{c_2} t^a)}{k_2^\pm + i\epsilon} + \frac{\tr(t^{c_2} t^{c_1} t^a)}{k^\pm_1 + i\epsilon}\right)n^\pm_{\nu_1} n^\pm_{\nu_2}  \notag \\
= {\bm q}^2 \bigg[ \frac{f^{a_1a_2c}}{2} &\bigg(  \frac{1}{k_1^\pm + i\epsilon} + \frac{1}{k_1^\pm - i\epsilon} \bigg)
 +  \frac{d^{a_1a_2c}}{2} \text{sgn}(\epsilon) 2\pi\delta(k^\pm_1)   
\bigg]
  n^\pm_{\nu_1} n^\pm_{\nu_2} ,
\end{align}
with the condition $k_2^\pm = -k_1^\pm$
implied. Eq.~\eqref{eq:indu_1L_eps} coincides with
eq.~\eqref{eq:indu1eps} up to the second term in the second line
proportional to the symmetric structure constant $d^{abc}$. The
symmetric color structure, the dependence on the sign of $\epsilon$
 and its symmetry behavior under $k_1^- \to -k_1^- $ indicate that
this term corresponds to a reggeized gluon with positive
signature. This is in contrast with the original formulation of the
effective action which contains only negative signatured reggeized
gluons.  While it  seems at first natural to use
eq.~\eqref{eq:indu_1L_eps} instead of eq.~\eqref{eq:indu1eps} and to
extend the effective action to reggeized gluons with positive
signature, we do not make use of this possibility in the following and
discard terms with symmetric color structure.  This choice is due to
the following reasons:

(i) The dependence on the sign of $\epsilon$ leads to a potential
conflict with factorization of QCD amplitudes in the high energy
limit. It denotes a dependence of the induced vertex on the energy of
the high energy factorized scattering parton (for an illustrative QCD
example we refer to Appendix \ref{sec:possig}).
From the point of view of the effective Lagrangian this dependence on
the sign of $\epsilon$ can be understood as a violation of hermicity
due to the simple replacement $D_\pm \to D_\pm -\epsilon$.

(ii) A further problem is connected with  symmetric color tensors such as
$d^{abc}$ itself. Unlike color tensors build from anti-symmetric
SU$(N_c)$ structure constants, such as the color tensors of the
unregulated induced vertices
figs.~\ref{fig:feynrules0p2}.c, \ref{fig:tra2}, \ref{fig:trans3}, the
symmetric tensors depend generally on the SU$(N_c)$ representation of
the fields in the Lagrangian\footnote{We are particular thankful to
  L.~N.~Lipatov for drawing our attention to this point.}. In the
case of QCD amplitudes this can be backtracked to the SU$(N_c)$
representation of scattering particles, signaling another potential
breakdown of  high energy factorization\footnote{That there exists
  indeed a dependence on the representation of scattering particles
  apart from normalization factors has been directly observed in
  \cite{Bartels:2009zc}, where an
  additional contribution for scattering particles in the adjoint
  representation has been found, which is not present for the
  fundamental representation.}.

In the following we therefore take the most conservative choice and
define the pole prescription of the induced vertices through the
replacement $D_\pm \to D_\pm -\epsilon$ together with a subsequent
projection on the 'maximal antisymmetric' sub-sector of its color
tensors, which is given in terms of anti-symmetric SU$(N_c)$ structure
constants alone.  The precise meaning of this projection onto the
maximal-anti-symmetric sector will be defined in short.  Leaving in
eq.~\eqref{eq:indu_1L_eps} aside the projector on the color octet
$\tr(\cdot t^a)$, where the dot represents any product of SU$(N_c)$
generators, we deal in case of the induced vertex of order $g$, with a
color tensor with two adjoint indices $c_1$ and $c_2$.  The maximal
anti-symmetric sub-sector of order two is then given by the commutator
of the two generators, while its symmetric counterpart is defined as
the anti-commutator\footnote{Note that without specifying the
  symmetric counterpart a projection on the anti-symmetric part is
  meaningless as the remainder is completely arbitrary and therefore
  also the resulting pole prescription. }. Projection on the maximal
anti-symmetric sub-sector corresponds therefore for the order $g$
induced vertex to dropping the symmetric term proportional to
$d^{abc}$ in eq.~\eqref{eq:indu_1L_eps}, which leaves us with the
commonly used pole prescription eq.~\eqref{eq:indu1eps}.

\subsection{Generalization to the order $g^2$ induced vertex}
\label{sec:defpole}

To generalize this projection to higher order induced vertices, we need
to find at first an appropriate  basis that generalizes the
decomposition in commutator and anti-commutator of the previous
subsection. This basis requires  two elements with a
double-commutator, which then yield the color structure of the induced
vertex fig.~\ref{fig:tra2}. In the following we use a short-cut
notation
\begin{align}
  \label{eq:com}
[1,2]&  = [t^{c_1},t^{c_2} ], & S_n (1 \ldots n) &= \frac{1}{n!} \sum_{i_1, \ldots i_n}t^{c_{i_1}} \ldots  t^{c_{i_n}},
\end{align}
denoting the commutator of two generators with color indices $c_1$ and
$c_2$ and symmetrization of $n$ generators respectively, where in the second
expression the sum is taken over all permutations of the numbers $1,
\ldots, n$.  In this notation, a possible decomposition of a color
tensor with three adjoint indices is given by the following basis
  \begin{align}
  \label{eq:change3_symbolic}
[[3,1],2], \quad [[3,2],1],  
\quad S_2\left([1,2]3 \right) \quad S_2\left([1,3]2 \right) \quad S_2\left([2,3]1 \right)
 \quad S_3\left(123 \right).
\end{align}
As a  commutator of two SU$(N_c)$ generators can be expressed in terms of a single generator, expressions such as $S_2([1,3]2 ) $  are well defined {\it e.g.}
\begin{align}
  \label{eq:s132}
S_2\left([1,3]2 \right) = t^{a_1}t^{a_3}t^{a_2}  - t^{a_3}t^{a_1}t^{a_2} 
+ t^{a_2}t^{a_1}t^{a_3} - t^{a_2}t^{a_3}t^{a_1}.
\end{align}
The terms in  eq.~(\ref{eq:change3_symbolic}) contains  two
double-anti-symmetric, three mixed-symmetric and one totally symmetric
element, while the third double-antisymmetric element $[[1,2],3]$ can
be expressed in terms of $[[3,1],2] $, $[[3,2],1] $ by means of the
Jacobi-identity.  The two double-antisymmetric elements $[[3,1],2] $,
$[[3,2],1] $ define  a basis of the maximal anti-symmetric
subsector of order two.  The above decomposition shares some
properties with the usual Young-tableau decomposition, while the
definition of anti-symmetrization differs in the present
case. Apparently the elements $[[3,1],2] $, $[[3,2],1] $ define  a
basis of the maximal anti-symmetric sub-sector of a color tensor with
three adjoint indices.  To obtain the pole prescription of the induced
vertex of order $g^2$, we start from the effective action with the
following expression
\begin{align}
 \label{eq:indu_2L_epsa}
   g^2v_\pm& \frac{1}{\partial\pm -\epsilon} v_\pm \frac{1}{\partial_\pm -\epsilon} v_\pm\partial^2_{\sigma} A_\mp,
\end{align}
which is obtained through the replacement $D_\pm \to D_\pm - \epsilon$
in eq.~\eqref{eq:1efflagrangian} and subsequent expansion in $g$. On
the level of Feynman rules, it results into the following unprojected
induced vertex of order $g^2$
\begin{align}
 \label{eq:indu_2L_eps}
 -ig^2 2{\bm q}^2
   \bigg( &
 \frac{\tr(t^{c_1} t^{c_2} t^{c_3} t^a)}{(k_2^\pm + k_3^\pm+ i\epsilon)(k_3^\pm + i\epsilon)} 
 +
\frac{\tr(t^{c_2} t^{c_1} t^{c_3} t^a)}{(k_1^\pm + k_3^\pm+ i\epsilon)(k_3^\pm + i\epsilon)}
\notag \\
&
+
\frac{\tr(t^{c_1} t^{c_3} t^{c_2} t^a)}{(k_2^\pm + k_3^\pm+ i\epsilon)(k_2^\pm + i\epsilon)} 
 +
\frac{\tr(t^{c_3} t^{c_1} t^{c_2} t^a)}{(k_1^\pm + k_2^\pm+ i\epsilon)(k_2^\pm + i\epsilon)}
\notag \\
&
 +
\frac{\tr(t^{c_3} t^{c_2} t^{c_1} t^a)}{(k_1^\pm + k_2^\pm+ i\epsilon)(k_1^\pm + i\epsilon)}
 +
\frac{\tr(t^{c_2} t^{c_3} t^{c_1} t^a)}{(k_1^\pm + k_3^\pm+ i\epsilon)(k_1^\pm + i\epsilon)}
\bigg) n^\pm_{\nu_1} n^\pm_{\nu_2}  n^\pm_{\nu_3}, 
\end{align}
with $k_1^\pm + k_2^\pm + k_3^\pm = 0$ implied. Leaving aside for the moment the factor
$-ig^2 2{\bm q}^2 n^\pm_{\nu_1} n^\pm_{\nu_2} n^\pm_{\nu_3} $ and also the
projection on the color octet, $\tr (\,\cdot \, t^a)$,    eq.(\ref{eq:indu_2L_eps}) reads in the basis eq.(\ref{eq:change3_symbolic})
\begin{align}
    -\frac{1}{6} \left[[3,2],1\right] \bigg(&
            \frac{2}{( k^\pm_3 -i\epsilon) ( k^\pm_1 +i\epsilon) } 
            +
            \frac{2}{( k^\pm_3 +i\epsilon) ( k^\pm_1 -i\epsilon) } 
              +
            \frac{1}{( k^\pm_2 +i\epsilon) ( k^\pm_3 -i\epsilon) } 
\notag \\            
  +&
            \frac{1}{( k^\pm_2 -i\epsilon) ( k^\pm_3 +i\epsilon) } 
            +
            \frac{1}{( k^\pm_1 +i\epsilon) ( k^\pm_2 -i\epsilon) } 
            +  
            \frac{1}{( k^\pm_1 -i\epsilon) ( k^\pm_2 +i\epsilon) } \bigg) \notag \\
 -\frac{1}{6} \left[[3,1],2\right]\bigg(&
            \frac{2}{( k^\pm_3 -i\epsilon) ( k^\pm_2 +i\epsilon) } 
            +
            \frac{2}{( k^\pm_3 +i\epsilon) ( k^\pm_2 -i\epsilon) } 
              +
            \frac{1}{( k^\pm_1 +i\epsilon) ( k^\pm_3 -i\epsilon) } \bigg)
\notag \\            
  +&
            \frac{1}{( k^\pm_1 -i\epsilon) ( k^\pm_3 +i\epsilon) } 
            +
            \frac{1}{( k^\pm_2 +i\epsilon) ( k^\pm_1 -i\epsilon) } 
            +  
            \frac{1}{( k^\pm_2 -i\epsilon) ( k^\pm_1 +i\epsilon) } \bigg) \notag \\
 -\frac{1}{2} S_2\left([1,2]3 \right)\bigg(&
            \frac{1}{( k^\pm_3 -i\epsilon) ( k^\pm_1 +i\epsilon) } 
            -
            \frac{1}{( k^\pm_3 +i\epsilon) ( k^\pm_1 -i\epsilon) } 
            +
            \frac{1}{( k^\pm_2 -i\epsilon) ( k^\pm_3 +i\epsilon) } 
\notag \\            
  -&
            \frac{1}{( k^\pm_2 +i\epsilon) ( k^\pm_3 -i\epsilon) } 
            +
            \frac{1}{( k^\pm_1 +i\epsilon) ( k^\pm_2 -i\epsilon) } 
            -  
            \frac{1}{( k^\pm_1 -i\epsilon) ( k^\pm_2 +i\epsilon) } \bigg) \notag \\
- \frac{1}{2} S_2\left([1,3]2 \right)\bigg(&
            \frac{1}{( k^\pm_2 -i\epsilon) ( k^\pm_1 +i\epsilon) } 
            -
            \frac{1}{( k^\pm_2 +i\epsilon) ( k^\pm_1 -i\epsilon) } 
            +
            \frac{1}{( k^\pm_3 -i\epsilon) ( k^\pm_2 +i\epsilon) } 
\notag \\            
  -&
            \frac{1}{( k^\pm_3 +i\epsilon) ( k^\pm_2 -i\epsilon) } 
            +
            \frac{1}{( k^\pm_1 +i\epsilon) ( k^\pm_3 -i\epsilon) } 
            -  
            \frac{1}{( k^\pm_1 -i\epsilon) ( k^\pm_3 +i\epsilon) } \bigg)
\notag \\ 
 -\frac{1}{2}S_2\left([2,3]1 \right)  \bigg(&
            \frac{1}{( k^\pm_3 -i\epsilon) ( k^\pm_1 +i\epsilon) } 
            -
            \frac{1}{( k^\pm_3 +i\epsilon) ( k^\pm_1 -i\epsilon) } 
            +
            \frac{1}{( k^\pm_2 +i\epsilon) ( k^\pm_3 -i\epsilon) } 
\notag \\            
 - &
            \frac{1}{( k^\pm_2 -i\epsilon) ( k^\pm_3 +i\epsilon) } 
            +
            \frac{1}{( k^\pm_1 -i\epsilon) ( k^\pm_2 +i\epsilon) } 
            -  
            \frac{1}{( k^\pm_1 +i\epsilon) ( k^\pm_2 -i\epsilon) } \bigg)
 \notag \\
 -S_3\left(123 \right)  \bigg(&
            \frac{1}{( k^\pm_3 -i\epsilon) ( k^\pm_1 +i\epsilon) } 
            +
            \frac{1}{( k^\pm_3 +i\epsilon) ( k^\pm_1 -i\epsilon) } 
            +
            \frac{1}{( k^\pm_2 -i\epsilon) ( k^\pm_3 +i\epsilon) } 
\notag \\            
  +&
            \frac{1}{( k^\pm_2 +i\epsilon) ( k^\pm_3 -i\epsilon) } 
            +
            \frac{1}{( k^\pm_1 +i\epsilon) ( k^\pm_2 -i\epsilon) } 
            +  
            \frac{1}{( k^\pm_1 -i\epsilon) ( k^\pm_2 +i\epsilon) } \bigg) .
 \label{eq:eikonal_inbasis}
\end{align}
Projection onto the maximal anti-symmetric sub-sector of order three
sets then all terms which contain an $S_2$ or an $S_3$ symbol to zero,
leaving only the color tensors contained in the unregulated vertex
fig.~\ref{fig:tra2}.  The pole structure can be further simplified
making use of the eikonal identity
\begin{align}
  \label{eq:eikonal_id}
\frac{1}{ k_1^\pm + i\epsilon} \frac{1}{ k_1^\pm + k_2^\pm + i\epsilon}  + \frac{1}{ k_2^\pm + i\epsilon} \frac{1}{ k_1^\pm + k_2^\pm + i\epsilon} = \frac{1}{ k_1^\pm + i\epsilon} \frac{1}{ k_2^\pm + i\epsilon},
\end{align}
which holds for the above pole prescription not only in the algebraic
sense but also in the sense of the theory of distributions
\cite{Bassetto:1991ue}.  Evaluating commutators and adding the
color-octet projection together with the common factor of
eq.(\ref{eq:indu_2L_eps}), which corresponds to the substitution
\begin{align}
  \label{eq:substution_symb_f}
 \left[[3,1],2\right]  & \to -ig^2{\bm q}^2 n^\pm_{\nu_1} n^\pm_{\nu_2} n^\pm_{\nu_3}f^{a_3a_2a}f^{a_1ac}, \notag \\
 \left[[3,2],1\right]& \to
-ig^2{\bm q}^2 n^\pm_{\nu_1} n^\pm_{\nu_2} n^\pm_{\nu_3}f^{a_3a_1a}f^{a_2ac} , 
\end{align}
we obtain for the pole prescription of the order $g^2$  induced vertex 
\begin{align}
  \label{eq:double_comm_simpli}
   \parbox{2.6cm}{\includegraphics[height = 2cm]{indu2.eps}}
   \begin{array}[h]{l } \\
\displaystyle
 =  -ig^2{\bm q}^2    \bigg[ f^{c_3c_2e}f^{c_1ea}
   g_2^\pm(3,1,2) \\  \displaystyle  \qquad \qquad \qquad 
          +  f^{c_3c_1e}f^{c_2ea}
       g_2^\pm(3,2,1)\bigg] n^\pm_{\nu_1} n^\pm_{\nu_2} n^\pm_{\nu_3} ,     
   \end{array}
\end{align}
where
\begin{align}
  \label{eq:g2}
g_2^\pm(i,j, m)=  \bigg[&
            \frac{-1/3}{ k^\pm_i\! -\!i\epsilon } 
            \bigg( \frac{1}{  k^\pm_m \!+\!i\epsilon} 
                                       + \frac{1/2}{  k^\pm_m\! -\!i\epsilon} \bigg) 
+
             \frac{-1/3}{ k^\pm_i \!+\!i\epsilon } 
            \bigg( \frac{1}{  k^\pm_m \!-\!i\epsilon} 
                                       + \frac{1/2}{  k^\pm_m \!+\!i\epsilon} \bigg) 
                             \bigg].
\end{align}
Eq.(\ref{eq:g2}) can be further compactified, making use of the identity,
\begin{align}
  \label{eq:delta_func}
\frac{1}{k^\pm + i\epsilon} - \frac{1}{k^\pm -i\epsilon} = -2\pi i \delta(k^\pm),
\end{align}
leading to 
\begin{align}
  \label{eq:cpv_rep}
g_2^\pm(i,j,m) = 
    \bigg[&  \frac{-1}{[k_i^\pm][k_m^\pm]} -\frac{\pi^2}{3}\delta(k_i^\pm)\delta(k_m^\pm) \bigg].
\end{align}
Using the eikonal identity for Cauchy principal values
eq.~\eqref{eq:eik_id_cpv}  and the condition $k_1^\pm + k_2^\pm + k_3^\pm =
0$ we find that the eikonal function $g_2$ obeys itself the eikonal
identity in the purely algebraic sense {\it i.e.}
\begin{align}
  \label{eq:identity_poles}
g^\pm_2(3,2,1) = - g_2(1,3,2) - g_2(3,1,2).
\end{align}
Bose-symmetry of eq.~(\ref{eq:cpv_rep}) is then easily verified.
Invariance of eq.~\eqref{eq:double_comm_simpli} under $\epsilon \to -
\epsilon$ and symmetry under $\{ k_1^\pm, k_2^\pm, k_3^\pm \} \to\{
-k_1^\pm, -k_2^\pm, -k_3^\pm \} $ in analogy to the unregulated vertex
fig.~\ref{fig:tra2} are also satisfied by the regulated vertex.

A comment is in order concerning the two delta functions appearing in
eq.~\eqref{eq:cpv_rep}. At first sight they seem to be in conflict
with the high energy expansion of underlying QCD amplitudes which
requires non-zero $k_i^\pm$, $i = 1,2,3$. These delta-functions appear
however only due to the identity eq.~\eqref{eq:delta_func} which
allows to reduce eq.~\eqref{eq:g2} into the more compact expression
eq.~\eqref{eq:cpv_rep}.  Eq.~\eqref{eq:g2} and the corresponding
expressions in eq.~\eqref{eq:eikonal_inbasis} are on the other hand
free of any delta-functions and can be therefore directly related to
an expansion of QCD Feynman diagrams. In addition one should note that
any product of two Cauchy principal values can be related to a product
of two delta-functions through the identity eq.~\eqref{eq:eik_id_cpv}
and therefore demanding complete absence of such terms is hard to
achieve.

\subsection{Pole prescription of higher order induced vertices}
\label{sec:indu3}
To define the pole prescription of the order $g^3$ and higher vertices
we follow the same pattern. The basis of the color tensor with four
adjoint indices $t^{c_1} t^{c_2}t^{c_3} t^{c_4}$ has now 24
independent elements. Again it can be decomposed into a six
dimensional maximal anti-symmetric sub-sector which has a basis in
terms of combined anti-symmetric structure constants alone and
symmetrization of lower dimensional maximal anti-symmetric
sub-sectors. A possible decomposition is given by
 \begin{align}
  \label{eq:change4_symbolic}
  \begin{array}[h]{c|c|c|c|c}
\left[\left[[4,1 ],2  \right],3   \right] &
S_2\left([1,2],[3,4]\right) &
S_2\left([[1,2] ,3]4\right) &
S_3\left([1,2]34\right) &
S_4\left(1234\right) \\
\left[\left[[4,1 ],3  \right],2   \right] &
S_2\left([1,3],[2,4]\right) &
S_2\left([[3,2] ,1]4\right) &
S_3\left([1,3]24\right)  &
 \\
\left[\left[[4,2 ],1  \right],3   \right] &
S_2\left([1,4],[2,3]\right)  &
S_2\left([[3,2] ,1]4\right)  &
S_3\left([1,4]23\right) &
 \\
\left[\left[[4,2 ],3  \right],1   \right] &
&
S_2\left([[1,2] ,4]3\right) &
S_3\left([2,3]14\right) &  \\
\left[\left[[4,3 ],2  \right],1   \right] &
&
S_2\left([[4,2] ,1]3\right) &
S_3\left([2,3]14\right) &
 \\
\left[\left[[4,3 ],1  \right],2   \right] &
&
S_2\left([[1,3] ,4]2\right) &
S_3\left([2,4]13 \right) &
\\
 &
&
S_2\left([[4,3] ,1]2\right) &
S_3\left([3,4]12 \right) &
\\ &&
S_2\left([[2,3] ,4]1\right) && \\ &&
S_2\left([[4,3] ,2]1\right) &&\\
  \end{array}
\notag\\ &.
\end{align}
To determine the pole prescription of the order $g^3$ induced vertex we start from the  following expression at Lagrangian level
\begin{align}
  \label{eq:indu3_starteff}
 -g^3v_\pm& \frac{1}{\partial_\pm -\epsilon} v_\pm \frac{1}{\partial_\pm -\epsilon} v_\pm \frac{1}{\partial_\pm -\epsilon} v_\pm\partial^2_{\sigma} A_\mp,
\end{align}
which is obtained from the replacement $D_\pm \to D_\pm - \epsilon$
and subsequent expansion in $g$.  Re-writing the color tensors of the
resulting vertex in terms of the above basis and setting all tensors
to zero, apart from the basis elements of the maximal anti-symmetric
sector of order four, contained in the first column of
eq.~\eqref{eq:change4_symbolic}, we obtain the following regulated
order $g^3$ induced vertex
\begin{align}
  \label{eq:pole_indu3}
& \parbox{3cm}{\includegraphics[height = 2cm]{indu3.eps}}  = 
-g^3 {\bm q}^2
n^\pm_{\nu_1} n^\pm_{\nu_2} n^\pm_{\nu_3} n^\pm_{\nu_4} \, \cdot  \notag \\
\bigg[&
 f^{a_4a_1d_2}f^{d_2a_3d_1}f^{d_1a_2c} 
 g_3^\pm(4,1,3,2)
+
f^{a_4a_1d_2}f^{d_2a_2d_1}f^{d_1a_3c}
 g_3^\pm(4,1,2,3)
\notag \\
+&
f^{a_4a_2d_2}f^{d_2a_1d_1}f^{d_1a_3c}
 g_3^\pm(4,2,1,3)
+
f^{a_4a_2d_2}f^{d_2a_3d_1}f^{d_1a_1c}
 g_3^\pm(4,2,3,1)
\notag \\
+&
f^{a_4a_3d_2}f^{d_2a_1d_1}f^{d_1a_2c}
 g_3^\pm(4,3,1,2)
+
f^{a_4a_3d_2}f^{d_2a_2d_1}f^{d_1a_1c}
 g^\pm_3(4,3,2,1) 
\bigg],
\end{align}
where the function $g_3^\pm(i,j,m,n)$ is defined as
\begin{align}
\label{eq:functionf}
& g_3^\pm  (i,j,m,n)  =\frac{(-1)}{12}\bigg\{ \frac{1}{k^\pm_i + i\epsilon} \bigg[
               \frac{1}{k_n^\pm  +  i\epsilon} 
\left( \frac{1}{k_n^\pm  +  k_m^\pm  +  i\epsilon }  +   \frac{1}{k_n^\pm + k_m^\pm - i\epsilon }\right)  +  
\notag \\
&               \frac{1}{k_n^\pm  - i\epsilon}  \left(   
               \frac{3}{k_n^\pm  +  k_m^\pm  - i\epsilon }
               +
               \frac{1}{k_n^\pm + k_m^\pm  + i\epsilon }
\right)
                                           \bigg]
 +  \frac{1}{k_i^\pm -i\epsilon} \bigg[
               \frac{1}{k_n^\pm  -  i\epsilon}
 \bigg(  \frac{1}{k_n^\pm + k_m^\pm - i\epsilon }  \notag \\
&  +   \frac{1}{k_n^\pm +  k_m^\pm  + i\epsilon }\bigg) 
  +  
\frac{1}{k_n^\pm  + i\epsilon} \left(   
               \frac{3}{k_n^\pm  +  k_m^\pm  + i\epsilon }
               +
               \frac{1}{k_n^\pm  + k_m^\pm  - i\epsilon }
\right)
                                           \bigg]
\bigg\}.
\end{align}
Similarly to eq.~\eqref{eq:g2} this function $g_3$ can be written in terms of  Cauchy-principal values and delta-functions, 
\begin{align}
  \label{eq:f_cpv}
g_3^\pm(i,j,m,n)
 =\bigg(&
          \frac{-1}{[k_i^\pm][k_n^\pm + k_m^\pm][k_n^\pm]}
            -
            \frac{\pi^2}{3} \delta(k_n^\pm) \delta(k_m^\pm) \frac{-1}{[k_i^\pm]} 
             \notag \\
             &
-
            \frac{\pi^2}{3} \delta(k_n^\pm) \delta(k_i^\pm) \frac{1}{[ k_m^\pm]} 
        - 
          \frac{\pi^2}{3} \delta(k_n^\pm + k_m^\pm) \delta(k_i^\pm) \frac{1}{[k_n^\pm]} \bigg).
\end{align}
Using eq.(\ref{eq:eik_id_cpv}),  eikonal identities such as
\begin{align}
  \label{eq:ids_indu4}
g_3^\pm(4,1,3,2) &= g_3^\pm(2,3,1,4) +g_3^\pm(2,3,4,1)+g^\pm_3(k_2^-, k_1^-,  k_3^-, k_4^-),
\end{align}
can be shown to hold, which allow to prove Bose-symmetry of the
induced vertex with the above pole-prescription.  Invariance under
$\epsilon \to - \epsilon$ and anti-symmetry under the substitution $
\{ k_1^\pm, k_2^\pm. k_3^\pm, k_4^\pm \} \to \{-k_1^\pm,- k_2^\pm
-k_3^\pm, -k_4^\pm \}$ in accordance with the behavior of the
unregulated vertex fig.~\ref{fig:trans3} can be shown to hold 
for the regulated vertex eq.~\eqref{eq:pole_indu3}.
At this stage the general recipe for the construction of the pole
prescription of the induced vertices should be clear.  The starting
point is the projector $P_{\text{A}}^{(1)}$ which acts trivially on
the single generator,
\begin{align}
  \label{eq:proj1}
   P_{\text{A}}^{(1)} t^{a_1} & = t^{a_1},
\end{align}
and defines in this way the maximal  antisymmetric sector of order one. 
To arrive at the projector $ P_{\text{A}}^{(n)}$ which projects the
color tensor with $n$ adjoint indices $t^{a_1} .... t^{a_n}$ on its
maximal anti-symmetric sub-sector we proceed as follows. We first
construct a basis for all possible color tensors with $n$ adjoint
indices which can be obtained through symmetrization of maximal
anti-symmetric sub-sectors of order $n-1$ or lower. The color tensors
with $n$ adjoint indices which are orthogonal to this (partial)
symmetric sub-space, define then the maximal anti-symmetric sub-sector
of order $n$.  $ P_{\text{A}}^{(n)}$ then projects a generic color
tensor with $n$ adjoint indices onto this maximal anti-symmetric
sub-sector of order $n$.  We have explicitly verified up to $n=5$ that
this sub-sector has a basis in terms of $(n-1)!$ independent
combinations of anti-symmetric SU$(N_c$) structure constants, as
contained in the unregulated induced vertices.
Given  this definition of projectors, the pole prescription can be
given directly at Lagrangian level, if desired. This requires
to replace the unregulated operator $W_\pm(v)$ in the induced
Lagrangian eq.~\eqref{eq:1efflagrangian} by
\begin{align}
  \label{eq:Wreg}
  W^\epsilon_\pm [v] & =  \frac{1}{2} \left[ \mathcal{P}_{\text{A}} \left( v_\pm \frac{1}{D_\pm - \epsilon} \partial_\pm \right)  + \mathcal{P}_{\text{A}} \left( v_\pm \frac{1}{D_\pm + \epsilon} \partial_\pm \right) \right], 
\end{align}
where the symmetrization in $\epsilon$ has no effect on the resulting
Feynman rules and merely serves to ensure hermicity of the regulated
Lagrangian.  The projector $\mathcal{P}_{\text{A}}$ acts order by
order in $g$ on the SU$(N_c)$ color structure of the gluonic fields
$v_\pm(x) = - it^a v^a (x)$, 
\begin{align}
  \label{eq:Wreg2}
  \mathcal{P}_{\text{A}} \bigg(& v_\pm \frac{1}{D_\pm - \epsilon} \partial_\pm \bigg)  \equiv
 -i\bigg(  P^{(1)}_{\text{A}}(t^a)  v^a_\pm  - (-ig) v^{a_1}_\pm \frac{1}{\partial_\pm - \epsilon}  v^{a_2}_\pm  P^{(2)}_{\text{A}}\left(t^{a_1}t^{a_2} \right)
\notag \\
& \qquad 
 +   (-ig)^2 v^{a_1}_\pm \frac{1}{\partial_\pm - \epsilon}  v^{a_2}_\pm  \frac{1}{\partial_\pm - \epsilon}  v^{a_3}_\pm     P^{(3)}_{\text{A}}\left(t^{a_1}t^{a_2}t^{a_3} \right)  - \ldots  \bigg),
\end{align}
where $P_{\text{A}}^{(n)}$ are the projectors of the color tensors
with $n$ adjoint indices on the maximal anti-symmetric sub-sector of
order $n$.

\section{Summary and conclusion}
\label{sec:concl}

We derived a pole prescription for the higher order
induced vertices of Lipatov's high energy effective action which
respects the symmetry properties of the unregulated induced vertices
and leads to identical color structure for regulated and unregulated
vertices. Explicit expressions have been derived up to the order $g^3$
induced vertex, while a recipe for the determination of the
prescription of the order $g^n$ induced vertices $n \ge 4$ has been
proposed. The prescription has the additional advantage that it can be
related to an expansion of QCD diagrams in the high energy
limit. Future applications of these vertices are numerous: they will
be used for the determination of the 2-loop corrections to the gluon
Regge trajectory from the effective and the determination of NLO
corrections to effective vertices.  In addition they are highly
relevant to construct evolution equations and production vertices in
the presence of multiple reggeized gluon exchanges which are currently
investigated within the context of the effective action.

\subsubsection*{Acknowledgments}

I am deeply indebted for intense and numerous discussion to J.~Bartels
and L.~N.~Lipatov.  I also want to thank Agustin Sabio Vera for useful
comments.  Financial support from the German Academic Exchange Service
(DAAD), the MICINN under grant FPA2010-17747, the Research Executive
Agency (REA) of the European Union under the Grant Agreement number
PITN-GA-2010-264564 (LHCPhenoNet) and the DFG graduate school
``Zuk\"unftige Entwicklungen in der Teilchenphysik'' is gratefully
acknowledged.

 \appendix 

 \section{Pole prescription and high energy limit of QCD}
 \label{sec:possig}

 In the following we take a closer look on the relation between the
 proposed pole prescription and the high energy limit of QCD Feynman
 diagrams. To start with,  we consider the high energy limit of the QCD $gq
 \to gq $ scattering at tree-level. Within QCD, this process is
 described by the sum of the following diagrams
\begin{align}
  \label{eq:gqqcd}
  \parbox{3cm}{\includegraphics[width = 3cm]{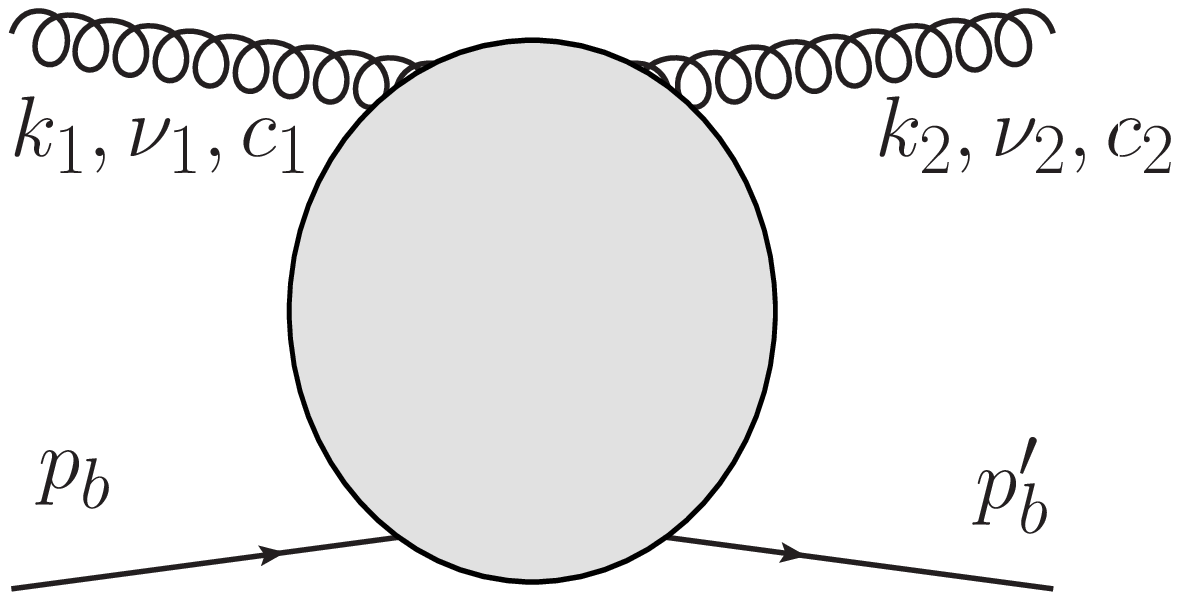}} & = 
 \parbox{3cm}{\includegraphics[width = 3cm]{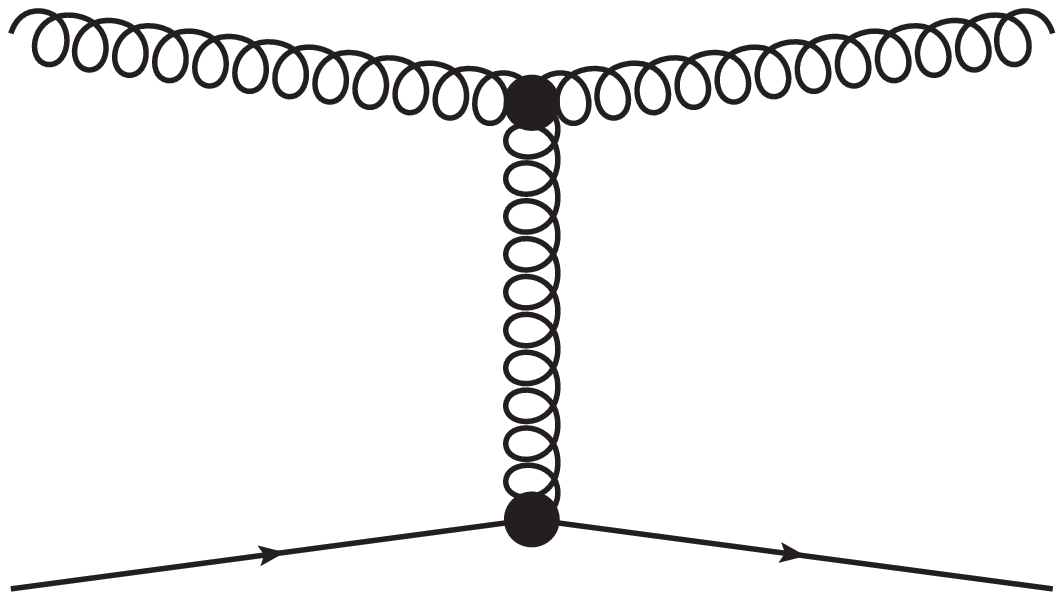}} + 
\parbox{3cm}{\includegraphics[width = 3cm]{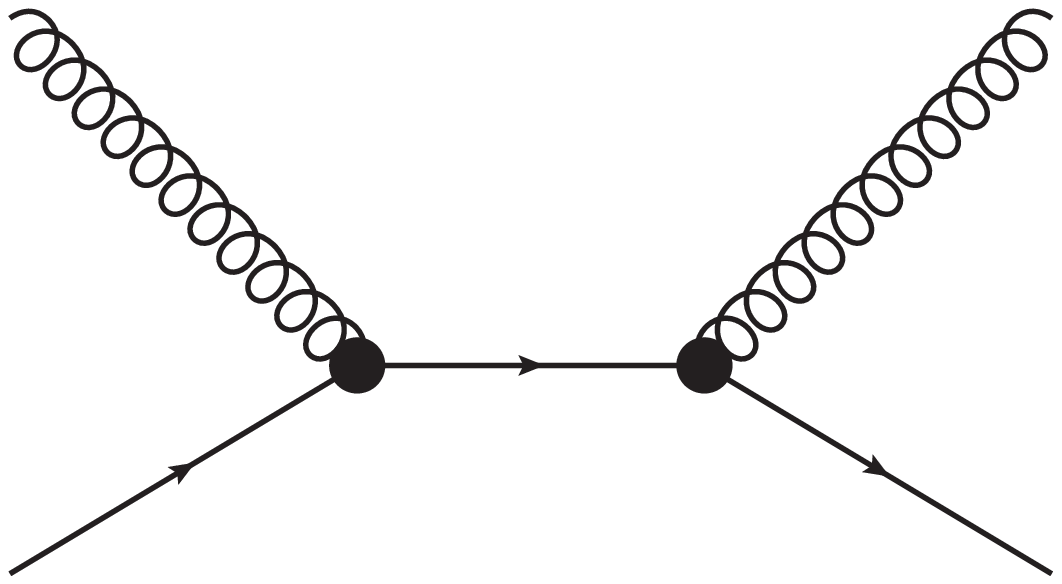}} + 
\parbox{3cm}{\includegraphics[width = 3cm]{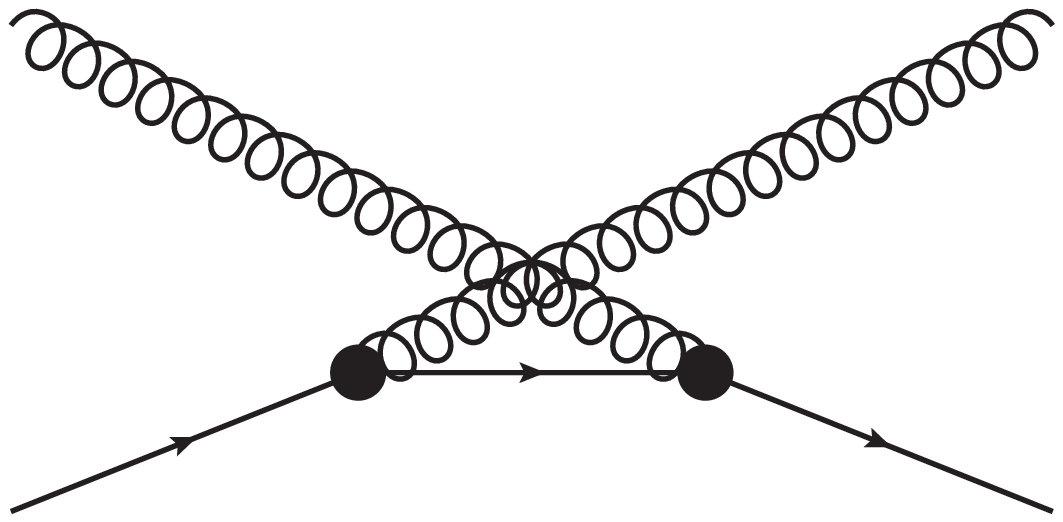}}.
\end{align}
The effective action describes on the other hand the same process (up to corrections suppressed by powers of $s = 2 p_a \cdot p_b$) as
\begin{align}
  \label{eq:gqeff}
  \parbox{3cm}{\includegraphics[width = 3cm]{gq_gen.eps}} & = 
 \parbox{3cm}{\includegraphics[width = 3cm]{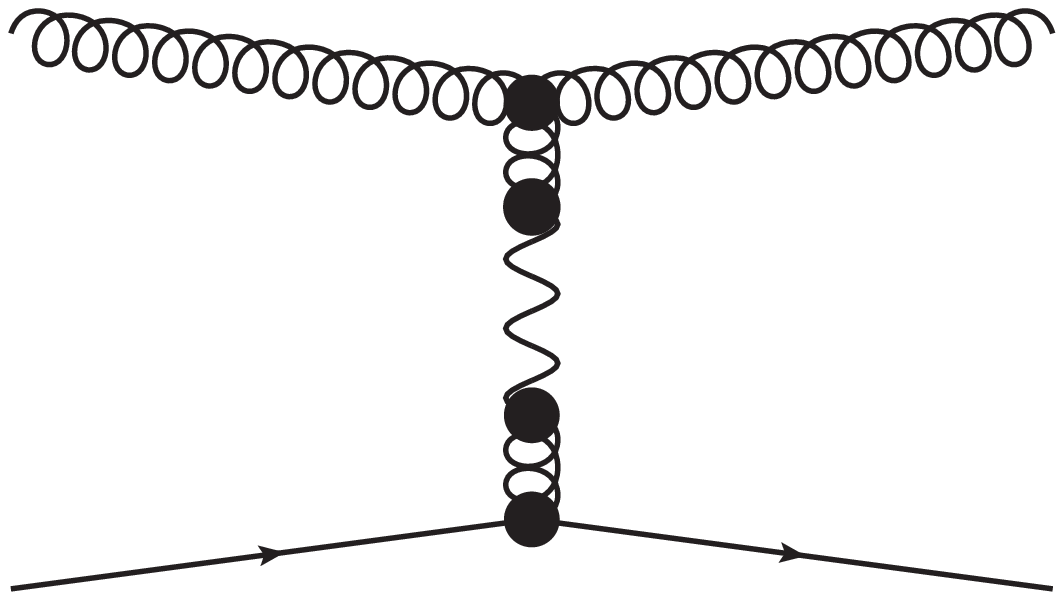}} + 
\parbox{3cm}{\includegraphics[width = 3cm]{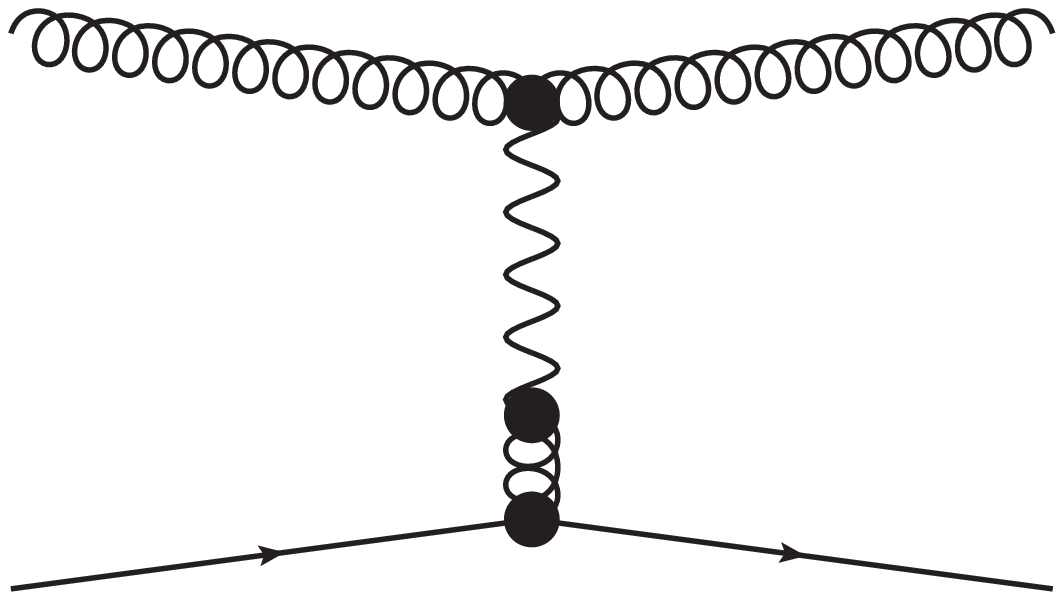}} + \mathcal{O}\left (\frac{|t|}{s} \right),
\end{align}
with $t = q^2 = (p_a - p_1)^2$.  For covariant gauges, to which we
restrict here, the first effective diagram can be identified with the
high energy limit of the first QCD diagram, while the second effective
diagram which contains the induced vertex of order $g$ can be
identified with the high energy limit of the second and third QCD
diagram. Leaving aside the polarization vectors of the gluons one has in the high energy limit
\begin{align}
  \label{eq:23qcd}
&   \parbox{3cm}{\includegraphics[width = 3cm]{gq_qcd2.eps}}  + 
\parbox{3cm}{\includegraphics[width = 3cm]{gq_qcd3.eps}}  =
\notag \\
&
 = \bar{u}(p_b') \left [  ig t^{c_2} \gamma^{\nu_2} \frac{i( \fdag{p}_b + \fdag{k}_1)}{(p_b + k_1)^2 + i\epsilon} ig t^{c_1} \gamma^{\nu_1} 
 +
 ig t^a \gamma^{\nu_1} \frac{i( \fdag{p}_b - \fdag{k}_2)}{(p_b - k_2)^2 + i\epsilon} ig t^c \gamma^{\nu_2}  \right] u(p_b)  \notag \\
&
= \bar{u}(p_b')   ig \fdag{n}^- \frac{i/2}{{\bm q}^2}  \left[
 \left(  \frac{ig  {\bm q}^2  t^{c_2} t^{c_1}    }{ k_1^+ + i\epsilon/p_b^-}  
-
 \frac{ig   {\bm q}^2 t^{c_1} t^{c_2}}{ k_1^+ - i\epsilon/p_b^-} \right) (n^+)^{\nu_1} (n^+)^{\nu_2}    \right]  u(p_b)  + \mathcal{O}\left( \frac{ |t|}{s} \right),
\end{align}
where the expression in the squared bracket can -- up to a projection
on the color octet sector -- be identified with the order $g$ induced
vertex as obtained in eq.~\eqref{eq:indu_1L_eps}. This is the core of
the statement that the replacement $D_\pm \to D_\pm - \epsilon$ leads
to a pole prescription in accordance with  the high energy expansion
of QCD Feynman diagrams.

While in the case of anti-symmetric color a dependence of the squared
bracket on the sign of $p_b^-$ is absent, the same is not true for the
symmetric color sector.  As a consequence high energy factorization of
eq.~\eqref{eq:23qcd} is only completely realized in the anti-symmetric
color sector. The latter changes the  sign under $s \to -s$ and
therefore corresponds -- from the point of view of Regge theory -- to
a negative signature. The sub-leading symmetric color sector, which is
invariant under $s \to -s$ and therefore carries positive signature,
keeps  a residual dependence on the light-cone energy
of the scattering quark and therefore does not factorize completely.  In
addition the resulting color tensor depends in this case on the
representation of the generators $t^d$ in eq.~\eqref{eq:23qcd}. Projecting on the color octet, we obtain for the symmetric sector  for generators in the
fundamental representation\footnote{We denote in this paragraph
  generators in the fundamental representation with a label F and
  generators in the adjoint representation with a label A. Apart from
  this paragraph $t^a$ always denotes a generator in the fundamental
  representation.} $t^a_F$ the symmetric SU$(N_c)$ structure constant
\begin{align}
  \label{eq:sum_fundi}
 \tr_F\left( \{t^a_F, t^b_F \} t^c_F\right) = \frac{1}{2} d^{abc}.
\end{align}
Generators in the  adjoint representation  $t^a_A$ yield instead a  zero result
\begin{align}
  \label{eq:sum_adj}
  \tr_A\left(\{t^a_A, t^b_A \}t^c_A\right) = 0.
\end{align}
For the anti-symmetric sector, the  commutator leads in both cases to the anti-symmetric structure constant $f^{abc}$. 
The above mapping from the prescription $D_\pm \to D_\pm - \epsilon$ to the
prescription obtained from Feynman diagrams holds also for higher
order induced vertices. For the order $g^2$ induced vertex one starts in this case  with the $gq \to ggq$ scattering
process in Quasi-Multi-Regge-Kinematics where the final state gluons
are of similar rapidity. The Feynman diagrams  which are  within Feynman gauge  associated with the order $g^2$ induced vertex read
\begin{align}
  \label{eq:gg}
  &  \parbox{3.2cm}{\includegraphics[width = 3cm]{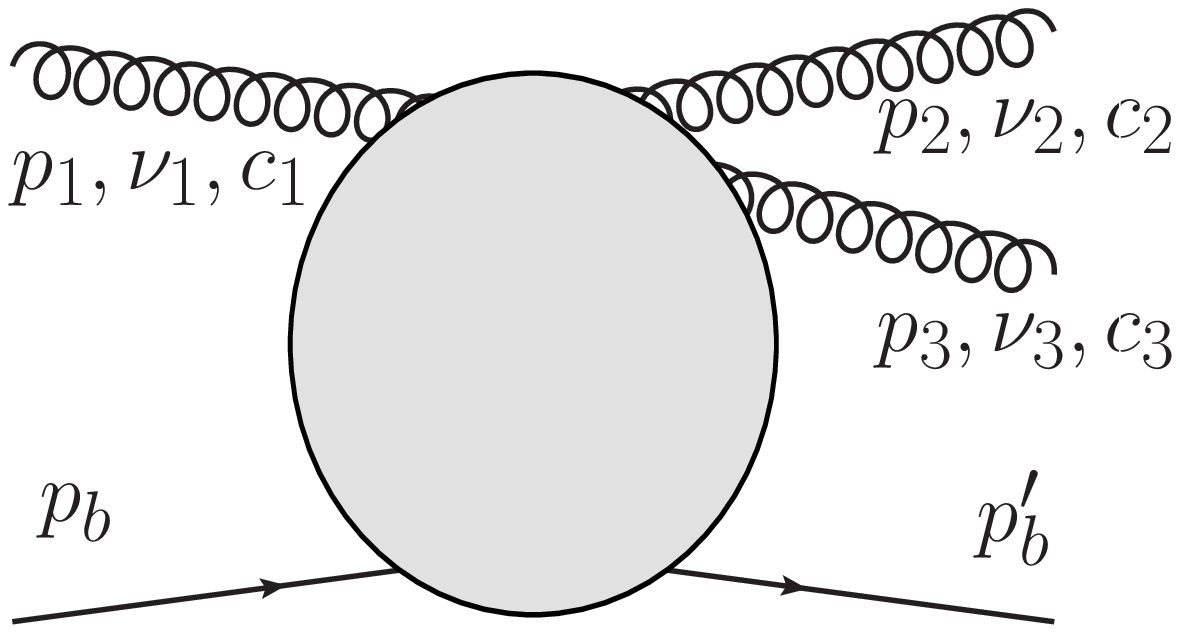}}  =  \parbox{3cm}{\includegraphics[width = 3cm]{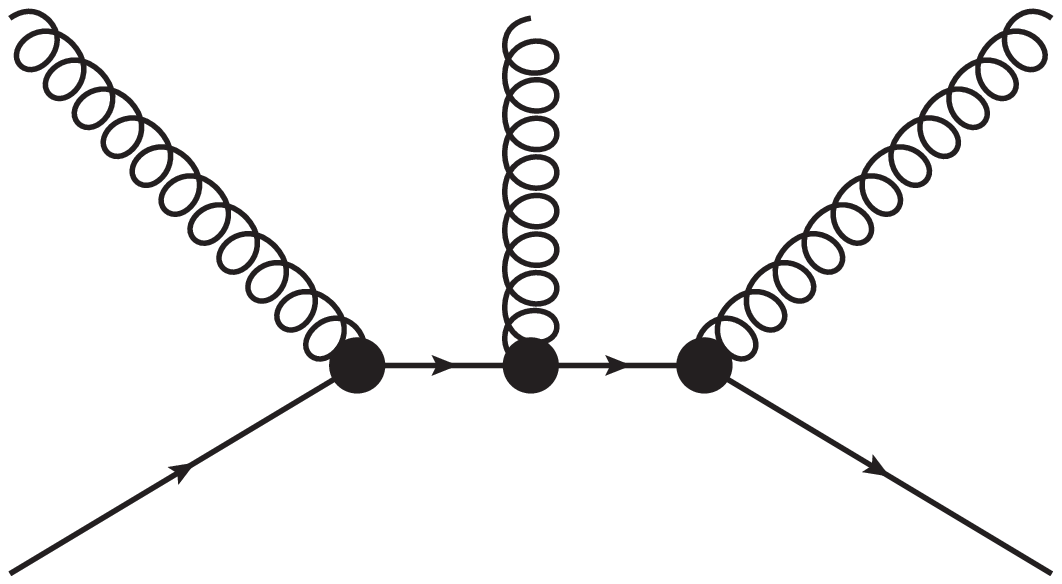}}\!\! + \!\! \parbox{3cm}{\includegraphics[width = 3cm]{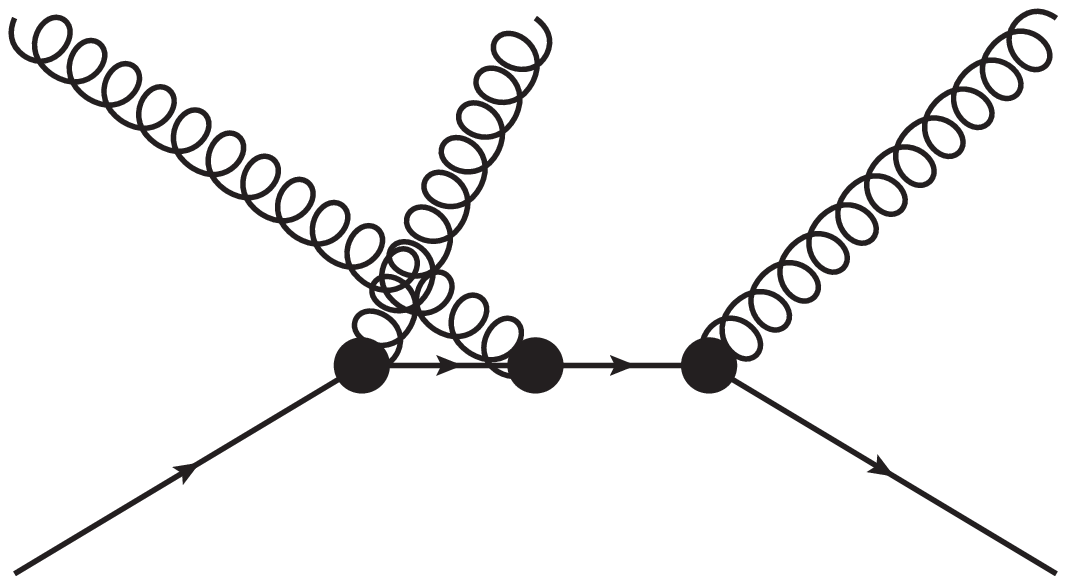}}  \!\!
+\! \! \parbox{3cm}{\includegraphics[width = 3cm]{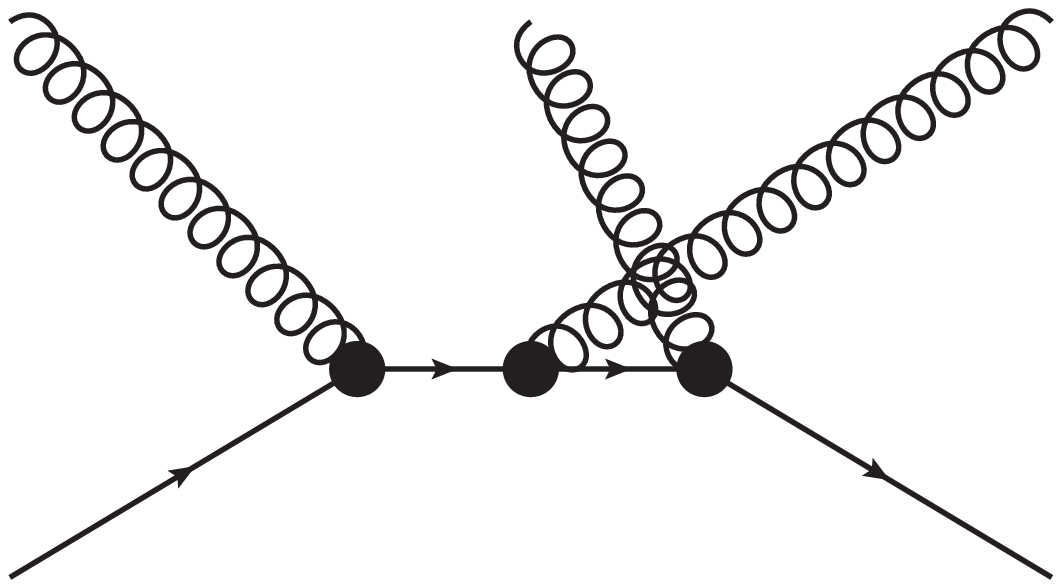}}  \notag \\
&  +  \parbox{3cm}{\includegraphics[width = 3cm]{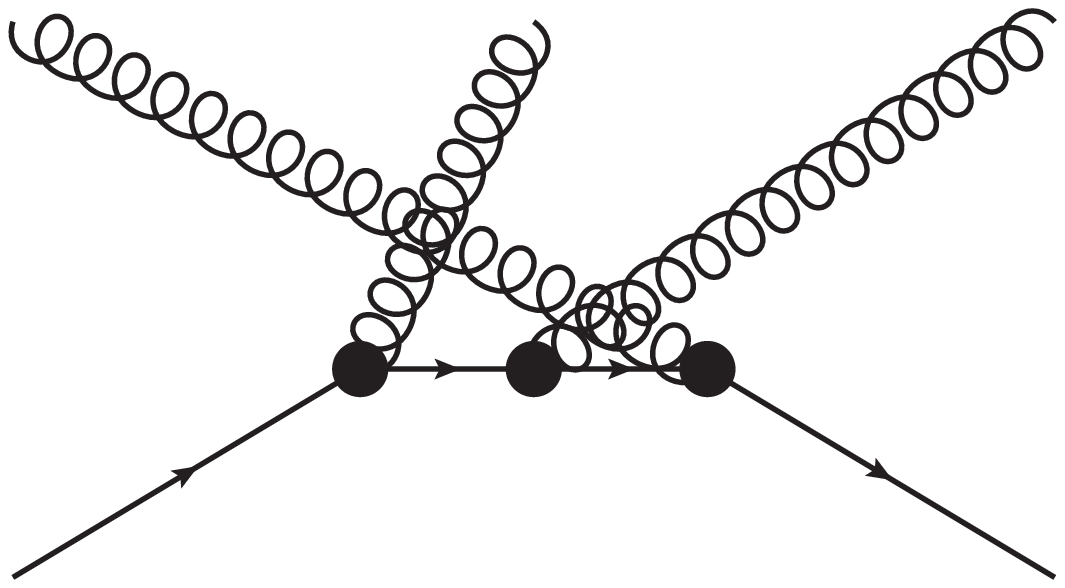}}    +  \parbox{3cm}{\includegraphics[width = 3cm]{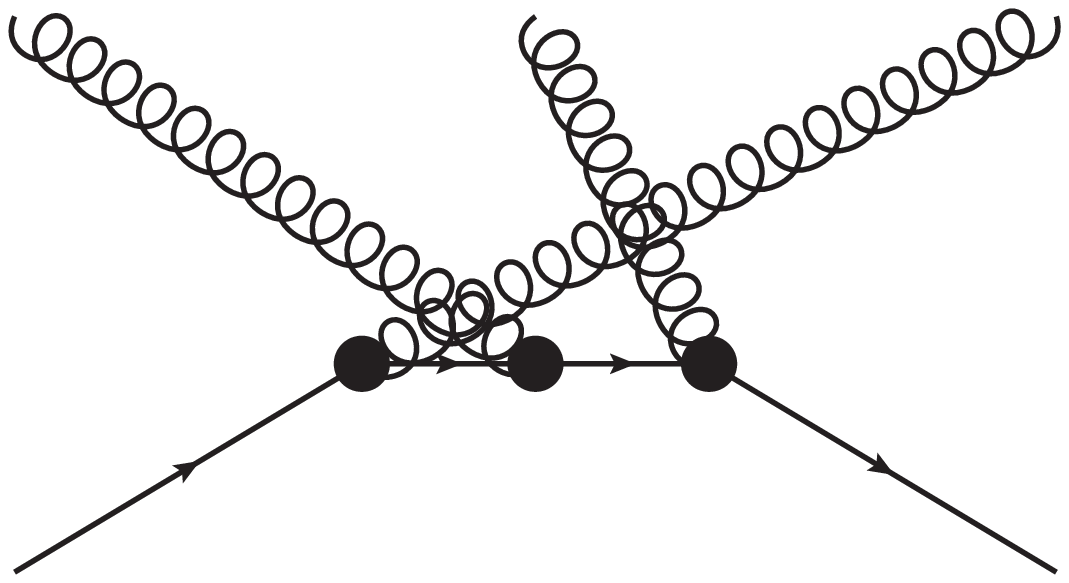}} +  \parbox{3cm}{\includegraphics[width = 3cm]{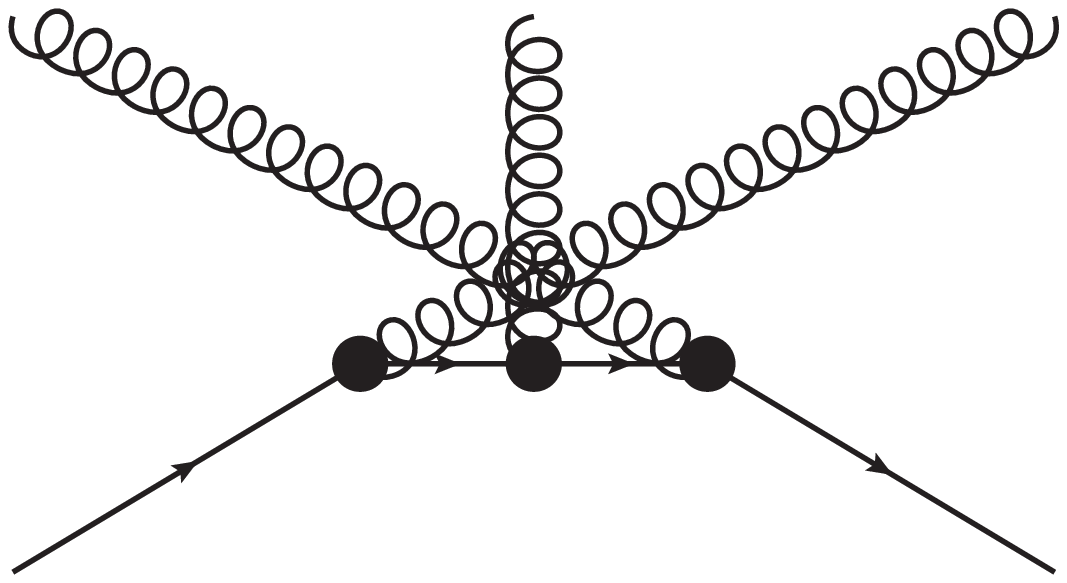}} + \, \ldots \, ,
\end{align}
where the dots indicates Feynman diagrams which can be associated
with combinations of lower induced vertices and pure QCD vertices.
Taking now all gluon momenta to be incoming, and leaving again aside
the polarization vectors of the gluons one finds for the sum of diagrams  depicted
diagrams in eq.~\eqref{eq:gg} in  the high energy limit
\begin{align}
  \label{eq:gghe}
& \bar{u}(p_b')   ig \fdag{n}^- u(p_b) \frac{i  /2}{{\bm q}^2}    \bigg[ 
 \bigg(  \frac{-ig^2  {\bm q}^2 t^{c_3} t^{c_2} t^{c_1}    }{( k_1^+ + i\epsilon/p_b^-)(- k_3^+ + i\epsilon/p_b^-)}  
+   \frac{-ig^2  {\bm q}^2 t^{c_3} t^{c_1} t^{c_2}    }{( k_2^+ + i\epsilon/p_b^-)(- k_3^+ + i\epsilon/p_b^-)}    
\notag \\   &  \qquad + 
 \frac{-ig^2 {\bm q}^2  t^{c_2} t^{c_3} t^{c_1}    }{( k_1^+ + i\epsilon/p_b^-)(- k_2^+ + i\epsilon/p_b^-)}   + 
 \frac{-ig^2 {\bm q}^2  t^{c_2} t^{c_1} t^{c_3}    }{( k_3^+ + i\epsilon/p_b^-)(- k_2^+ + i\epsilon/p_b^-)}  
\notag \\
 & \qquad \,   +
 \frac{-ig^2 {\bm q}^2  t^{c_1} t^{c_3} t^{c_2}    }{( k_2^+ + i\epsilon/p_b^-)(- k_1^+ + i\epsilon/p_b^-)}   + 
  \frac{-ig^2   {\bm q}^2  t^{c_1} t^{c_2} t^{c_3}    }{( k_3^+ + i\epsilon/p_b^-)(- k_1^+ + i\epsilon/p_b^-)}  
\bigg) \notag \\
& \qquad \qquad \qquad \qquad \qquad\qquad \qquad\qquad \qquad  \cdot
(n^+)^{\nu_1} (n^+)^{\nu_2}  (n^+)^{\nu_3}   \bigg]  + \mathcal{O}\left( \frac{{\bm q}^2}{s} \right),
\end{align}
from where the relation to expressions such as
eq.~\eqref{eq:indu_2L_eps} becomes clear. Similar results hold then
in an apparent way for higher order induced vertices.

\end{document}